\documentclass[5p,times,twocolumn]{elsarticle}
\usepackage{amsmath,amssymb,hyperref}
\usepackage[dvips]{epsfig}
\usepackage{slashed}
\usepackage{mathrsfs}
\usepackage{graphicx}
\usepackage{xcolor}
\usepackage{hyperref}
\usepackage{dcolumn}
\usepackage{subfigure}
\usepackage{xspace}
\usepackage{bbm}

\definecolor{heidelbeer}{rgb}{0.5,0,0.5}

\graphicspath{{./Figures/}
}
\begin{document}

\begin{frontmatter}

\title{The inverse problem for Schwinger pair production}

\author[be]{F. Hebenstreit}
\ead{hebenstreit@itp.unibe.ch}

\address[be]{Albert Einstein Center for Fundamental Physics, Institute for Theoretical Physics, Bern University, 3012 Bern, Switzerland}

\begin{abstract}
 The production of electron-positron pairs in time-dependent electric fields (Schwinger mechanism) depends non-linearly on the applied field profile.
 Accordingly, the resulting momentum spectrum is extremely sensitive to small variations of the field parameters.
 Owing to this non-linear dependence it is so far unpredictable how to choose a field configuration such that a predetermined momentum distribution is generated.
 We show that quantum kinetic theory along with optimal control theory can be used to approximately solve this inverse problem for Schwinger pair production.
 We exemplify this by studying the superposition of a small number of harmonic components resulting in predetermined signatures in the asymptotic momentum spectrum.
 In the long run, our results could facilitate the observation of this yet unobserved pair production mechanism in quantum electrodynamics by providing suggestions for tailored field configurations. 
\end{abstract}


\end{frontmatter}

\section{Introduction}  
The vacuum breakdown in external electric fields by the emission of the lightest charged particle-antiparticle pairs (Schwinger effect) is an unobserved prediction of quantum electrodynamics (QED) \cite{Sauter:1931zz,Heisenberg:1935qt,Schwinger:1951nm}.
Until recently, this pair production mechanism seemed to be a mere academic problem owing to the required electric field strength of the order of $E_S\sim10^{18}\,V/m$.
Recent efforts have raised the hopes that a direct observation might become feasible at ultra-high intensity laser facilities \cite{DiPiazza:2011tq,eli,xfel,xcels}.
Alternatively, there have been suggestions to study the Schwinger effect in graphene \cite{Allor:2007ei,Gavrilov:2012jk,Fillion-Gourdeau:2015dga} or to quantum simulate it in analogue cold atom systems \cite{Szpak:2011jj,Kasper:2015cca}. 

The pair production problem in QED cannot be solved in full glory but requires an approximate treatment.
Assuming ultra-high intensity lasers, it is well-justified to neglect quantum fluctuations of the electromagnetic field and to treat it as a classical, external one.
In addition, since the spatio-temporal laser scales are orders of magnitude different from the QED scale, which is set by the electron mass $m$, the focal region is typically modeled by a spatially homogeneous, time-dependent electric field.
In this case, quantum kinetic equations are especially well-suited to study the pair production problem numerically since they can be formulated as a coupled system of ordinary first-order differential equations \cite{Schmidt:1998vi,Bloch:1999eu}, which can be solved very efficiently.
For more complicated field configurations including spatial inhomogeneities or non-vanishing magnetic fields, more sophisticated approaches such as worldline methods \cite{Gies:2005bz,Dunne:2005sx,Kim:2007pm,Ilderton:2015lsa}, the Dirac-Heisenberg-Wigner formalism \cite{BialynickiBirula:1991tx,Hebenstreit:2011wk,Hebenstreit:2011cr,Blinne:2013via} or methods from real-time lattice gauge theory \cite{Su:2012,Hebenstreit:2013qxa,Hebenstreit:2013baa,Kasper:2014uaa} have been employed.

It has been proposed to lower the threshold for pair production by superimposing electric fields with different frequencies (dynamically-assisted Schwinger effect) \cite{Schutzhold:2008pz} and the resulting momentum spectra have been investigated.
Several studies revealed a very rich structure of the momentum distribution, owing to the fact that the pair-production problem depends non-linearly on the applied field profile. 
In fact, certain characteristic features of the momentum distribution were interpreted in terms of symmetries of the electric field and resonance conditions \cite{Hebenstreit:2009km,Dumlu:2010ua,Orthaber:2011cm,Akkermans:2011yn,Abdukerim:2013vsa,Kohlfurst:2013ura,Otto:2014ssa,Akal:2014eua,Panferov:2015yda}.
However, it is almost impossible to predict the variations in the momentum spectrum upon changing the field parameters {\it a priori}, i.\,e.~without actually solving the dynamic equations.
A rare exception are solitonic fields, for which particle production can be excluded at one predetermined momentum mode \cite{Kim:2011jw,Kim:2011sf}.

From an experimental point of view, we have to face the challenge that the number of produced electron-positron pairs is exponentially small and therefore hard to discriminate from background noise.
Moreover, spectrometers are only sensitive in a certain momentum window and characterized by a finite momentum resolution.
Accordingly, it would be most beneficial to generate electric field configurations such that the momentum distribution is peaked in a characteristic momentum window in order to enhance the detection probability. 
This situation is reminiscent of the inverse scattering problem in quantum mechanics where asymptotic scattering data is used to reconstruct the scattering potential \cite{Chadan:1989}.

In this manuscript, we present a method to solve the inverse problem for Schwinger pair production, i.\,e.~we determine the field parameters resulting in a predetermined momentum distribution, by combining quantum kinetic theory with optimal control theory.
We employ optimization techniques which have been utilized previously in performing pulse-shape optimization for pair production \cite{Kohlfurst:2012rb,Hebenstreit:2014lra}, and which have also proven successful in the closely related field of atomic, molecular and optical physics \cite{Chu:2001,Christov:2001,BenHajYedder:2004}.
Motivated by advances in high harmonic and attosecond pulse generation \cite{Brabec:2000zz,Goulielmakis:2004,Gordienko:2005zz,Sansone:2006}, we superimpose a small number of harmonic components and optimize their amplitudes such that a predetermined momentum distribution is well-approximated.
In the long run, our results could be used for tailoring superpositions of high harmonics in order to maximize the detection probability in a given momentum window.
On the other hand, a measured momentum distribution could also serve as a tomograph of the laser field, which is hard to control in an absolute manner at ultra-high intensities.


\section{Inverse problem for Schwinger pair production}
Vacuum electron-positron production in a unidirectional time-dependent electric field $E(t)=-\dot{A}(t)$ can be described in terms of a quasi-particle distribution function $F(q,t)$, where $q\equiv(q_\parallel,q_\perp)$ is the momentum variable with components along and perpendicular to the field direction, and $t$ denotes time.
Given a finite time interval $t\in[-T,T]$ with $E(\pm T)=0$, $F(q,T)$ can be regarded as the momentum distribution of asymptotic particles \cite{Dabrowski:2014ica}.
Defining auxiliary functions $G(q,t)$ and $H(q,t)$, its dynamics is governed by the coupled system of ordinary first-order differential equations \cite{Bloch:1999eu}
\begin{subequations}
\label{eq:eom}
\begin{alignat}{3}
 &\dot{F}&\ =\ &WG \ , \\
 &\dot{G}&\ =\ &W[1-F]-2\omega H \ , \\
 &\dot{H}&\ =\ &2\omega G \ ,
\end{alignat} 
\end{subequations}
with initial values $F(q,-T)=G(q,-T)=H(q,-T)=0$.
For $p_\parallel(t)=q_\parallel-eA(t)$ and $\epsilon_\perp^2=m^2+q_\perp^2$, we have $\omega^2(q,t)=\epsilon_\perp^2+p_\parallel^2(t)$ and $W(q,t)=eE(t)\epsilon_\perp/\omega^2(q,t)$.

In the following, we take a superposition of $n$ harmonic components on the time interval $t\in[-T,T]$
\begin{subequations}
\label{eq:fields}
\begin{align}
 E(t)&=\sum_{j=1}^{n}\mathcal{E}_j\sin\left(\frac{\pi(t+T)j}{2T}\right) \ , \\
 A(t)&=\sum_{j=1}^{n}\frac{2\mathcal{E}_jT}{j\pi}\left[\cos\left(\frac{\pi(t+T)j}{2T}\right)-1\right] \ ,
\end{align}
\end{subequations}
such that $E(\pm T)=0$ and $A(-T)=0$.
We then mean to select the field amplitudes $\mathbf{E}\equiv(\mathcal{E}_1,\ldots,\mathcal{E}_n)\in\mathbbm{R}^n$ such that a predetermined distribution function $F_0(q)$ is obtained at the final time $T$. 
While this search could be achieved by brute force for a very small number of harmonics, this approach becomes practically unfeasible for larger values of $n$.
Hence we set up an optimization problem to solve this inverse problem for Schwinger pair production by defining the cost functional as the least square deviation of $F(q,T)$ from the objective distribution $F_0(q)$:
\begin{equation}
 \label{eq:cost}
 \eta_n[\mathbf{E}]=\frac{\int{[dq] \left[F(q,T)-F_0(q)\right]^2}}{\int{[dq] F_0^2(q) } } \ .
\end{equation} 
The global minimum of the cost functional in the space $\mathbbm{R}^n\ni\mathbf{E}$ spanned by the field amplitudes is then defined as
\begin{equation}
 \tilde{\eta}_n\equiv\min_{\mathbf{E}\in\mathbbm{R}^n}{\eta_n[\mathbf{E}]} \ .
\end{equation} 
For reasons which will become clear later, the cost functional will also be denoted as quality factor in the following. 
We emphasize that the distribution function $F(q,T)$ implicitly depends on the field amplitudes $\mathbf{E}=(\mathcal{E}_1,\ldots,\mathcal{E}_n)$ via the dynamical system \eqref{eq:eom}.
The constrained optimization problem can be recast in an unconstrained optimization problem by introducing Lagrange multiplier functions $\lambda_{F,G,H}(q,t)$ \cite{Kohlfurst:2012rb}, which need to obey the adjoint equations
\begin{subequations}
\label{eq:adj}
\begin{alignat}{3}
 &\dot{\lambda}_F&\ =\ &W\lambda_G \ , \\
 &\dot{\lambda}_G&\ =\ &-W\lambda_F-2\omega \lambda_H \ , \\
 &\dot{\lambda}_H&\ =\ &2\omega \lambda_G \ , 
\end{alignat}
\end{subequations} 
with final values $\lambda_G(q,T)=\lambda_H(q,T)=0$ and $\lambda_F(q,T)=2[F_0(q)-F(q,T)]/\int{[dq]F_0^2(q)}$.
The stationary condition $0=\nabla \eta_n[\mathbf{E}^*]\in\mathbbm{R}^n$ is then a necessary condition for being a local extremum, with
\begin{align}
 \label{eq:grad}
 \nabla &\eta_n[\mathbf{E}] = e\int{[dq]}\int_{-T}^{T}dt\left[\frac{2p_\parallel}{\omega}\left(\lambda_HG-\lambda_GH\right)\nabla A\right. \nonumber \\
            &\left.+\frac{\epsilon_\perp}{\omega^2}\left(\lambda_G F-\lambda_F G-\lambda_G\right)\left(\nabla E+\frac{2eEp_\parallel}{\omega^2}\nabla A\right)\right]  \ . 
\end{align}
Here, $\nabla\equiv(\partial_{\mathcal{E}_1},\ldots,\partial_{\mathcal{E}_n})$ is the gradient with respect to the field amplitudes $\mathcal{E}_j$, and $\mathbf{E}^*$ denotes a local minimizer configuration. 
Moreover, all occurring functions are supposed to be solutions of the dynamical equations \eqref{eq:eom} and \eqref{eq:adj}, respectively, and the field gradients $\nabla E, \nabla A\in\mathbbm{R}^n$ are easily determined from \eqref{eq:fields}.

To take full advantage of the gradient information \eqref{eq:grad}, we employ a multi-start method along with a local optimization algorithm.
To this end, we generate random trial configurations $\mathbf{E}_0$ which are sampled from a probability distribution in field amplitude space.
The corresponding local minimizer configurations $\mathbf{E}^*=\lim_{k\to\infty}\mathbf{E}_k$ are then found iteratively 
\begin{equation}
 \mathbf{E}_{k+1}=\mathbf{E}_k+\alpha_k\mathbf{d}_k \ , 
\end{equation}
with $k\in\mathbbm{N}^0$.
We calculate the search directions $\mathbf{d}_k$ according to the Broyden-Fletcher-Goldfarb-Shanno (BFGS) algorithm, and viable step sizes $\alpha_k$ are found via an inexact line search fulfilling the strong Wolfe conditions.
For further algorithmic details we refer to, e.~g.~reference \cite{Nocedal:2006}. 

As a word of caution we emphasize that it is only guaranteed for infinite sample sizes that the global minimum $\tilde{\eta}_n\equiv\eta_n[\tilde{\mathbf{E}}]$, where $\tilde{\mathbf{E}}$ denotes the global minimizer configuration, is among the detected stationary points \cite{Marti:2003}.
Moreover, it is also not guaranteed that an exact solution of the inverse problem, which is characterized by a vanishing cost functional $\tilde{\eta}_n=0$, actually exists.
In general, we will rather find $\tilde{F}(q,T)\neq F_0(q)$ for certain momenta, where $\tilde{F}(q,T)$ denotes the momentum distribution corresponding to the global minimum $\tilde{\eta}_n$.
This is in fact a sign that the inverse problem for Schwinger pair production is in general ill-posed \cite{Kabanikhin:2011}.
A configuration that approaches the exact solution to the inverse problem is characterized by $\tilde{\eta}_n\ll1$, and we will study its dependence on the number of harmonic components $n$ in the following.

\begin{figure}[t]
 \includegraphics[width=\columnwidth]{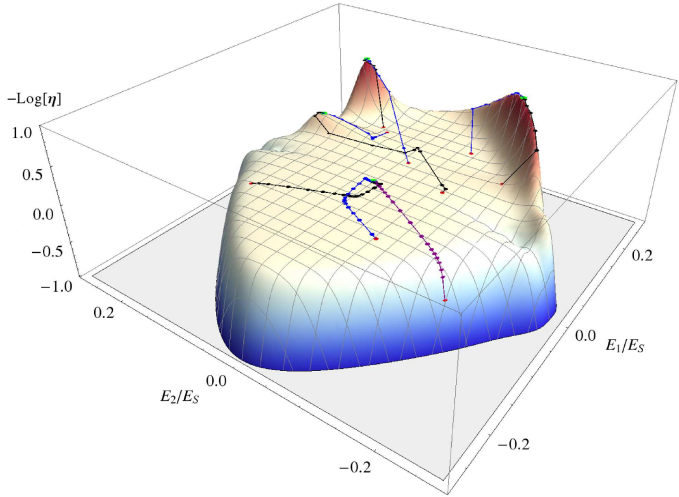}
 \caption{[Color online] 3D-plot of the quality factor landscape for a superposition of $n=2$ harmonics.
 For display reasons we show $-\log\eta_2[\mathcal{E}_1,\mathcal{E}_2]$, so that local maxima in the plot actually correspond to local minima of $\eta_2[\mathcal{E}_1,\mathcal{E}_2]$.
 In green we display local minimum configurations $\mathbf{E}^*$, amongst them the global minimum configuration $\tilde{\mathbf{E}}=(0.1644E_S,-0.1238E_S)$ with $\tilde{\eta}_2=0.1592$.
 We also display some typical optimization trajectories which lead from randomly chosen initial configurations ({\it red points}) to the local extrema ({\it green points}).}
 \label{fig1:potential}
\end{figure}

\section{Solving the inverse problem}
For simplicity, we study the inverse pair production problem in a $1$-dimensional system but emphasize that the general procedure is basically not altered in higher dimensions.
To be specific, we predefine a Gaussian model objective distribution 
\begin{equation}
 F_0(q)=A_0\exp\left(-\frac{(q-q_0)^2}{2\sigma^2_0}\right) \ ,
\end{equation}
and we take as parameters $A_0=10^{-5}$, $q_0=5m$ and $\sigma_0=m$.
We could specify any functional form in principle, however, we will restrict ourselves to this simple and instructive example.
Specifically, we consider the harmonic superposition \eqref{eq:fields} in the time interval $2T=100/m$. 

We first consider the superposition of $n=2$ harmonics and calculate the landscape $\eta_2[\mathcal{E}_1,\mathcal{E}_2]$ by brute force, cf.~Fig.~\ref{fig1:potential}.
We note that the distribution function $F_2(q,T)$ is computed for an array of field amplitudes $\mathcal{E}_i\in\{-0.275E_S,\ldots,0.275E_S\}$ with $\Delta\mathcal{E}_i=0.005E_S$, resulting in $\mathcal{O}(10^4)$ simulated data points.
We therefore conclude that the brute force approach becomes practically unfeasible for large values of $n$.\footnote{Given the chosen time extent $T$ and requiring sufficient momentum resolution, it takes a few minutes on a standard desktop computer to calculate the momentum distribution $F_2(q,T)$ according to \eqref{eq:eom}. 
Consequently, the computation time on the chosen array of field amplitudes is of the order of $T_2\sim \mathcal{O}(10^3)$ CPU hours. 
Increasing the number of harmonics $n>2$ and keeping the resolution $\Delta\mathcal{E}_i$ unaltered, the computation time grows exponentially $T_n\sim 100^{(n-2)}T_2$, indicating that the brute force approach becomes practically unfeasible for large values of $n$.}
The landscape shows a diamond-shaped region with quality factors $\eta_2[\mathbf{E}]=\mathcal{O}(1)$ which is rather flat, whereas it increases exponentially outside.
On the other hand, we also find a small number of distinct local minima which are indicated in green in Fig.~\ref{fig1:potential}, amongst them the global minimum $\tilde{\mathbf{E}}=(\tilde{\mathcal{E}}_1,\tilde{\mathcal{E}}_2)=(0.1644E_S,-0.1238E_S)$ with $\tilde{\eta}_2=0.1592>0$.
We emphasize that there is only an approximate solution to the inverse problem as the quality factor is non-vanishing. 

\begin{figure}[t]
 \includegraphics[width=\columnwidth]{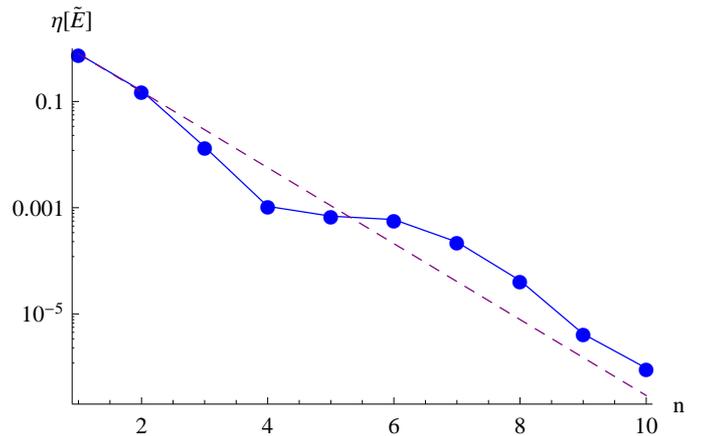}
 \caption{[Color online] Quality factor $\tilde{\eta}_n$ for the global minimizer configuration $\tilde{\mathbf{E}}$ as a function of the number of harmonics $n$ ({\it blue dots}). 
 The exponential fit $\tilde{\eta}_n\sim 4.1(2) \times \exp\left(-1.64(4)n\right)$ describes the trend reasonably well ({\it purple line}).}
 \label{fig2:quality}
\end{figure}

To illustrate the functionality of the employed optimization method, we first perform the numerical optimization for $n=2$ and we display some of the calculated trajectories in Fig.~\ref{fig1:potential}.
Depending on the randomly chosen initial configuration $\mathbf{E}_0$, the various trajectories converge to the diverse stationary points in the landscape $\eta_2[\mathcal{E}_1^*,\mathcal{E}_2^*]$.
For $n=2$, the average number of required iteration steps for sufficient convergence -- based on $N_0=200$ randomly chosen initial configuration -- is $\bar{k}_2=17.5$.

\begin{figure}[t!]
 \includegraphics[width=\columnwidth]{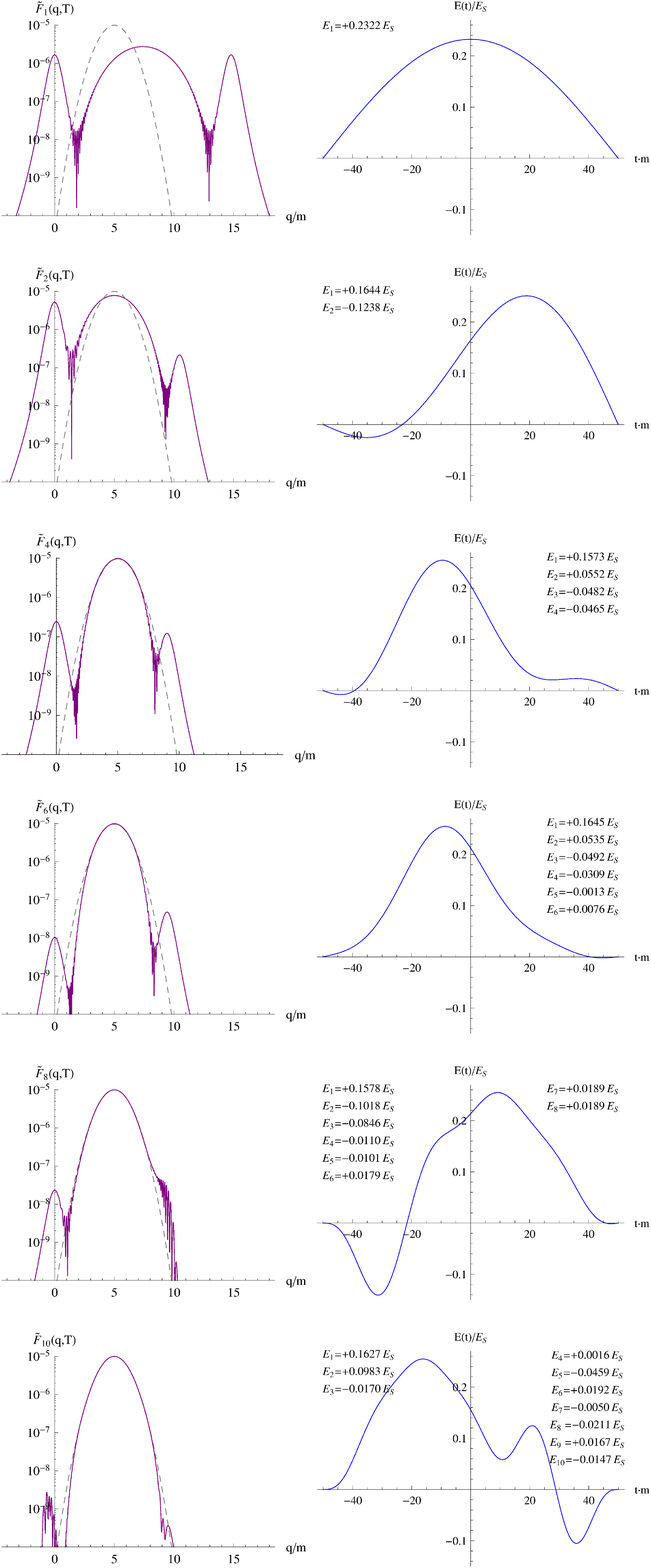} 
\caption{[Color online] {\it Left:} Logarithmic plot of the numerically determined global minimum distribution $\tilde{F}_{n}(q,T)$ for an increasing number of harmonics $n=1,2,4,6,8,10$. 
 {\it Right:} Corresponding electric field configuration $E(t)/E_S$.}
 \label{fig3:optimization_dist}
\end{figure}

In order to converge towards the objective momentum distribution $F_0(q)$, i.\,e.~to further decrease the quality factor $\tilde{\eta}_n$, we need to increase the number of harmonic components $n$.
Most notably, we then observe quick and substantial improvement of $\tilde{\eta}_n$.
This assertion is quantified in Fig.~\ref{fig2:quality}, where we display the quality factor $\tilde{\eta}_n$ as a function of $n$.
Moreover, we display the underlying global minimum distributions $\tilde{F}_{n}(q,T)$ for different values of $n$ along with the corresponding electric field configurations $E(t)$ in Fig.~\ref{fig3:optimization_dist}.
Based on our result, we note that the average number of required iteration steps grows like $\bar{k}_n\sim n$, indicating that the simulations become more expensive for increasing $n$. 
We emphasize, however, that this linear growth of the optimization approach outweighs by far the exponential growth of the brute force approach.

For $n=1$ we find an optimal distribution function $\tilde{F}_{1}(q,T)$ which exhibits a broad central peak accompanied by two narrow side peaks.
Notably, the central peak is located around $q_{1}=7.4m$ and is thus substantially shifted from $q_0=5m$.
Including one additional harmonic such that $n=2$, the optimal distribution function $\tilde{F}_{2}(q,T)$ becomes peaked around $q_{2}=5m=q_0$ but is still somewhat broader and smaller than $F_0(q)$.
Notably, we still observe two undesired peaks around $q=0$ and $q=10.5m$, respectively.
Further increasing the number of harmonics $n>2$, the optimal distribution $\tilde{F}_n(q,T)$ approximates the objective distribution $F_0(q)$ better and better.
Here, we emphasize that the global minimum configuration $\tilde{F}_n(q,T)$ does not necessarily arise from the global minimum configurations $\tilde{F}_{n-1}(q,T)$, meaning that $(\tilde{\mathbf{E}}_{n-1},0)\in\mathbbm{R}^n$
does not necessarily lie in the basin of attraction of $\tilde{\mathbf{E}}_{n}\in\mathbbm{R}^n$.
For instance, given the $n=2$ global minimum configuration $\tilde{\mathbf{E}}_2=(0.1644E_S,-0.1238E_S)$ with $\tilde{\eta}_2=\eta_2[\tilde{\mathbf{E}}_2]=0.1592$ (cf.~Fig.\ref{fig1:potential}) and taking these field parameters as the initial point for the $n=3$ optimization problem,  we converge towards the stationary point $\mathbf{E}^*_3=(0.1459E_S,-0.1436E_S,-0.0170E_S)$ with $\eta_3[\mathbf{E}^*_3]=0.1392$.
On the other hand, starting from the second-best $n=2$ stationary configuration $\mathbf{E}^*_2=(0.1561E_S,0.1314E_S)$ with $\eta_2[\mathbf{E}^*_2]=0.1604$ , we actually converge towards the $n=3$ global minimum configuration $\tilde{\mathbf{E}}_3=(0.1687E_S,-0.0345E_S,-0.0821E_S)$  with $\tilde{\eta}_3=[\tilde{\mathbf{E}}_3]=0.1399\cdot 10^{-1}$.
Consequently, the shape of the electric field can change quite substantially for different values of $n$ (cf.~Fig.~\ref{fig3:optimization_dist}).

The presented results indicate that a rather small number of harmonic components $n$ suffices to well-approximate specific predetermined features of the momentum distribution.
For instance, the distribution around $q_0=5m$ is already extremely well-described for $n=4$ whereas there are still comparatively large deviations at other momentum values present.
On the other hand, the number of harmonics $n$ needs to be further increased in order to approach the exact solution of the inverse problem for Schwinger pair production.
Quantitatively, we find that the quality factor drops from $\tilde{\eta}_1=0.7916$ to $\tilde{\eta}_{10} = 0.9525\cdot10^{-6}\ll1$.
Based on our results, the functional dependence of the quality factor is reasonably well-described by an exponential function $\tilde{\eta}_n\sim 4.1(2) \times \exp\left(-1.64(4)n\right)$.
One may therefore speculate that there is an exact solution of the inverse problem for the chosen objective distribution $F_0(q)$.
It is, however, not clear whether this solution then exists for finite $n$ or only in the limit $n\to\infty$.

We already mentioned that the employed multi-start method is guaranteed to converge to the global minimum only for infinite sample sizes.
Accordingly, Fig.~\ref{fig2:quality} is to be considered as an upper limit whereas there are, in principle, even better configurations conceivable.
In the current study, we have used $N_0=200$ random initial configurations $\mathbf{E}_0$ for all considered values of $n$.
Given this ensemble size, all the numerically determined global minimum distribution $\tilde{F}_n(q,T)$ were encountered several times. 
We are therefore confident that the identified configurations actually correspond to the global minimum $\tilde{\eta}_n$ or lie, at least, in its close neighborhood.

\section{Conclusions}
We demonstrated that quantum kinetic theory along with optimal control theory can be utilized to solve the inverse problem for Schwinger pair production.
Based on the instructive example of a Gaussian objective distribution $F_0(q)$, we found that it suffices to superimpose a comparatively small number of harmonics $n=4$ in order to well-approximate specific predetermined features.
These findings could substantially facilitate the observation of Schwinger pair production by providing suggestions for tailored field configurations.
In the long run, the presented method could also serve as a tomograph for applied laser pulses by reconstructing the applied field after measuring the asymptotic momentum distribution.
One has to be aware, however, that this procedure might be non-unique due to the ill-posedness of the inverse problem.
On the other hand, it is not clear at this point whether an exact solution of the inverse problem with $\tilde{\eta}_n=0$ actually exists and, if so, a finite number of harmonic components $n$ suffices. 
This issue is beyond the scope of the current investigation but should be further studied in the future.

The used optimization method is especially well-suited for parameter spaces which are not too high-dimensional.
In fact, we found that the average number of required iterations grows like $\bar{k}_n\sim n$, indicating that the optimization algorithm becomes less efficient for increasing $n$.
Moreover, the ensemble size $N_0$ of random initial configuration $\mathbf{E}_0$ needs to be enlarged upon further increasing the number of harmonics $n$, indicating that the multi-start approach becomes less efficient as well.
Accordingly, we should also envisage to apply global optimization strategies based on metaheuristic algorithms in the future in order to study the inverse problem for more intricate objective distributions $F_0(q)$, to investigate the possibility of an exact solution of the inverse problem with $\tilde{\eta}=0$, or to perform elaborate pulse shaping investigations.

\section*{Acknowledgments}
We like to thank F.~Fillion-Gourdeau for fruitful discussions and valuable comments, as well as R.~Alkofer, C.~Kohlf\"urst, M.~Mitter and G.~von~Winckel for collaboration on related work.
The research leading to these results has received funding from the European Research Council under the European Union's Seventh Framework Programme (FP7/2007-2013)/ ERC grant agreement 339220.

\end{document}